\documentclass[10pt]{article}
\usepackage{longtable}
\usepackage{tabu}

\usepackage[hyphens]{url}
\usepackage{scrextend}
\usepackage[hidelinks]{hyperref}

\usepackage{graphicx}
\graphicspath{ {./images/} }
\usepackage[ruled,linesnumbered]{algorithm2e}
\usepackage[numbers]{natbib}
\usepackage[singlelinecheck=false]{caption}
\DeclareCaptionLabelFormat{blank}{}

\usepackage{booktabs} 
\usepackage{authblk}
\usepackage{adjustbox}
\usepackage{color, colortbl}
\usepackage{caption}
\usepackage{subcaption} 
\usepackage{amsthm}
\usepackage{amsmath}

\usepackage{pdflscape}
\usepackage{enumitem}
\usepackage{lineno}
\usepackage{float}
\hypersetup{
    colorlinks=true,
    linkcolor=blue,
    citecolor=blue,
    urlcolor=blue}

\definecolor{Gray}{gray}{0.9}
\definecolor{White}{rgb}{1,1,1}
\definecolor{LightCyan}{rgb}{0.88,1,1}
\definecolor{orange}{rgb}{1,0.5,0}

\usepackage[space]{grffile}

\makeatletter
\g@addto@macro{\thm@space@setup}{\thm@headpunct{:}}
\makeatother

\makeatletter
\newcommand{\printfnsymbol}[1]{%
  \textsuperscript{\@fnsymbol{#1}}%
}
\makeatother

\begin{document}
\title{Ethnic Representation Analysis of Commercial Movie Posters}

\author[1]{Dima Kagan\thanks{kagandi@post.bgu.ac.il}\thanks{Authors contributed equally}}
\author[2]{Mor Levy\thanks{morxl1601@gmail.com}\printfnsymbol{2}}
\author[1]{Michael Fire\thanks{mickyfi@post.bgu.ac.il}}
\author[1]{Galit Fuhrmann Alpert\thanks{fuhrmann@bgu.ac.il}}
\affil[1]{Department of Software and Information Systems Engineering, Ben-Gurion University of the Negev, Israel}
\affil[2]{Afeka College of Engineering, Israel}

\maketitle

\begin{abstract}
In the last decades, global awareness towards the importance of diverse representation has been increasing.
Lack of diversity and discrimination toward minorities did not skip the film industry.
Here, we examine ethnic bias in the film industry through commercial posters, the industry's primary advertisement medium for decades.
Movie posters are designed to establish the viewer's initial impression.
We developed a novel approach for evaluating ethnic bias in the film industry by analyzing nearly 125,000 posters using state-of-the-art deep learning models.
Our analysis shows that while ethnic biases still exist, there is a trend of reduction of bias, as seen by several parameters.
Particularly in English-speaking movies, the ethnic distribution of characters on posters from the last couple of years is reaching numbers that are approaching the actual ethnic composition of US population.
An automatic approach to monitor ethnic diversity in the film industry, potentially integrated with financial value, may be of significant use for producers and policymakers.
\end{abstract}

\providecommand{\keywords}[1]{\textbf{Keywords:} #1}


\section{Introduction}
\label{chap:intro}


The film industry often portrays a Utopian image: Cocktail parties, glorious scenes shot behind magnificent views, and movie stars. This ideal image generates appreciation and worship of the general audience towards the industry and its stars. These affections are exactly those that have been harnessed towards effective and profitable advertising over the years \citep{kaikati1987celebrity,kamins1990investigation}.
Besides the obvious financial factors of movie advertisements, they also hold a critical role in shaping cultural and societal perception, as the film displays an image of society that viewers often perceive as being realistic \citep{carroll1985power}. 
This image is based upon many aspects, such as cast selection, key concepts, and visual stimulation, all of which may paint a picture that may be very different from our day to day reality.
Film content can influence people's actions in real-life, especially children, which are more vulnerable to such influence \citep{albert1957role,beaufort2019candy,Nieman2003ImpactOM}.
This shaping of mind may have a more significant impact than we realize, particularly with respect to diversity awareness.

With recent rise of the ``Black Lives Matter'' movement, diversity awareness has gained worldwide attention \citep{GeorgeFloyd, BlackLivesMatter}, highlighting the profound lack of ethnic diversity in one of the largest and most influential countries in the world, the USA. 
Recent reports indicate that similar issues are also evident specifically in the film industry \citep{HollywoodDiversityNewsroom, HollywoodDiversityMarketwatch}.
MarketWatch, for example, reported that the lack of non-White lead characters in Hollywood films is driven by economic concerns since, in some movie studios, having African-American leads is thought of as non-profitable when distributing films overseas \citep{HollywoodDiversityMarketwatch}.

Moreover, movie advertisements, such as posters and trailers,can reach more people than the actual film, exposing both viewers and non-viewers to the movie. As a consequence, movie posters have a particularly high impact on shaping society perception. 
In this study, we therefore aimed at quantifying diversity representation in the film industry, using movie posters.

\begin{figure*} 
         \centering
        \includegraphics[width=1\textwidth]{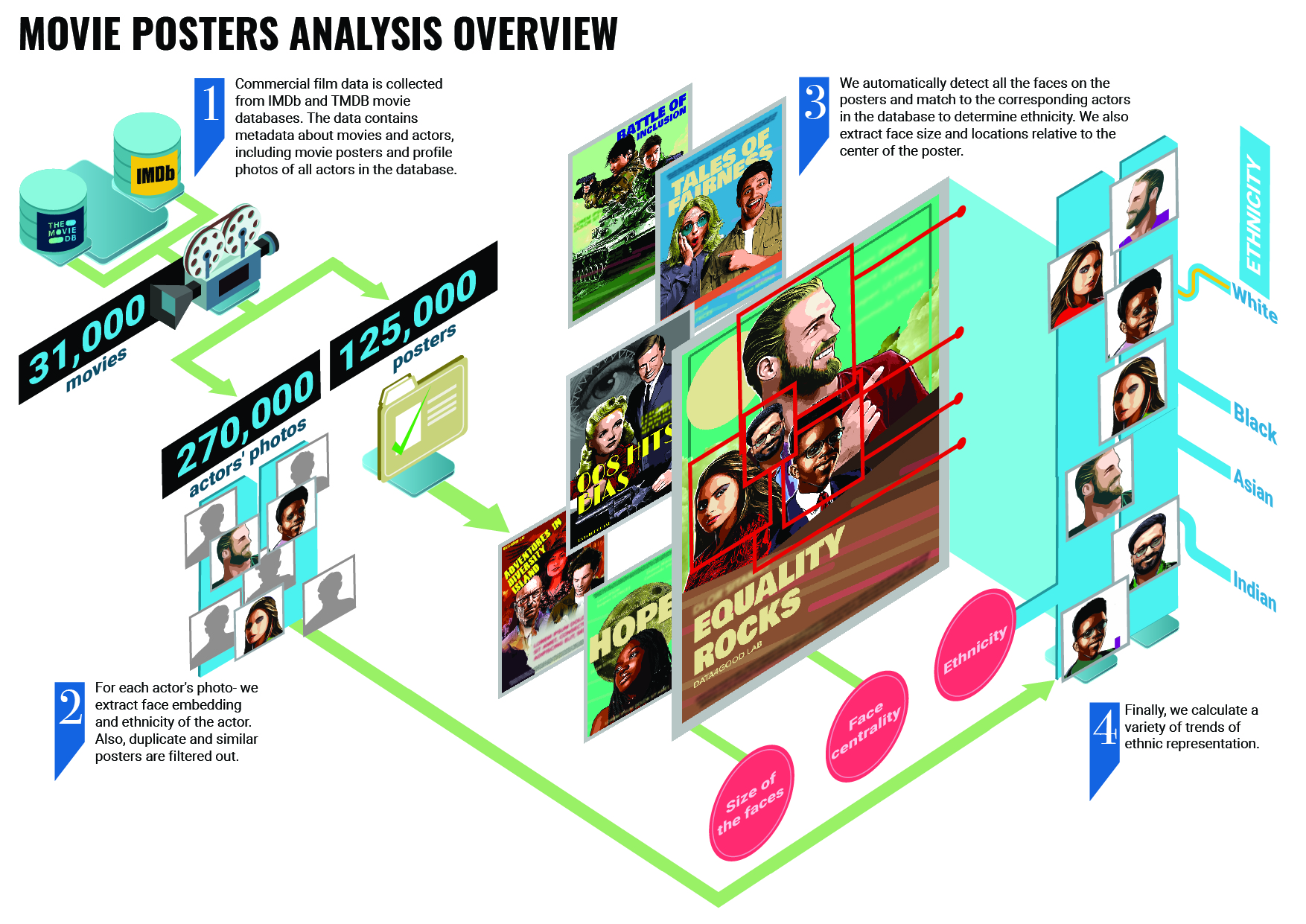}
        \caption{Movie posters analysis overview.}
        \label{fig:infog}
\end{figure*}

In this study, we developed a novel approach to automatically analyze and quantify diversity in movie posters (see Figure \ref{fig:infog}.)
Our posters dataset consists of nearly 125,000 unique posters that we collected from over 31,000 movies of various genres, nearly two thirds of them containing actors.
The data was collected from IMDb and TMDB online movie databases.
We used state-of-the-art machine learning based algorithms in order to identify patterns of interest that can shed light on ethnic diversity, in several steps, as follows.
For each poster, we first apply a face detection algorithm to recognize actors, followed by face embedding to match actors to their actual identity provided in IMDb and TMDB photos. 
We then apply deep learning to extract actors' ethnicity from their photos.
These analysis steps are used to generate a large-scale annotated poster 
dataset and an open-source code framework that we offer 
as publicly available resources for poster analysis.

The results of our large-scale analysis indicate that until recently White Caucasian actors were over-represented on posters compared to their relative numbers in the US populations, especially in leading roles. 
While in the last 20 years, a tendency for improvement in the representation of ethical minorities is observed, there are nevertheless over 79\% White Caucasian actors depicted on movie posters, regardless of the genre (see Figure \ref{fig:heatmap-genres-faces-races-count-all-years}), a fraction which is 1.14 higher than their relative demographic percentage in the total population.
We also found that non-White actors are more likely to be placed on a poster in minor roles.
Out of the top-3 actors from the cast list on posters, only 9.2\% where minorities.
In more minor roles (top 4-12), 16.2\% were non-White minority actors.

That being said, we will show that today the film industry is actually taking into account the ethnic diversity of the US population when creating posters. Posters of movies from the last two years are in fact perfectly balanced according to US population compound.
We propose that our approach could be used to continuously monitor ethnic representation in the film industry in an automated fashion.

The key contributions presented in this paper are five-fold:
\begin{itemize}
    \item We present a novel method that we developed for automatic evaluation of ethnic diversity.
    
    \item This is the first computer vision based study that uses poster images to analyze ethnic diversity.
    
    \item The described study is the most comprehensive study to date that utilizes posters to analyze ethnic representation in the film industry over decades.
    
    \item We curated and offered the largest publicly available poster dataset with 125,439 posters of 31,054 movies. The dataset consists of public metadata, as well as analyzed metadata that we automatically generated by face recognition and ethnicity detection of the identified actors faces in each poster.
    
    \item We provide an open-source analysis framework that we developed for movie posters analysis.
  Our framework could be used to generate metadata about movie posters, and facilitate research by creating and analyzing  larger amounts of data than has been available ever before.
    
    \end{itemize}

\section{Related Work}
\label{chap:rw}

The lack of diverse representation in the film industry is not unique to ethnic minorities. Similar under-representation may also be observed for other minority groups, including female actors, sexual minorities, religious groups, or various disability groups.

\subsection{Gender Diversity In the Film Industry}


Specifically, gender bias in the film industry is a well-studied concern. Studies have shown that women in the film industry are 
underrepresented, with an average male dominance of twice as many male speaking characters than female ones~\citep{lauzen2018sa, lauzen2018boxed,womenUnderrepresentedHollywood}.
Producers try to explain this financially; they claim that there is a greater profit for selling movie rights overseas for movies with internationally famous male actors ~\citep{fivethirtyeightWomenProducers}. 
Recent data science studies, using machine learning based approaches over large amounts of movie data, showed that female are indeed under-represented in movies~\citep{yang2020measuring} and women on average play more minor roles~\citep{kagan2019using}.

\subsection{Ethnic Under-Representation in Multiple Domains}
Ethnic under-representation is a general concern, not specific to the film industry alone.
It is present in multiple domains.
For example, \citet{koh2019offline} performed a large-scale empirical analysis aimed at detecting biases on Airbnb. Using image processing tools to automatically extract gender, age, and race information they showed that gender and racial homophily effects are apparent at all cities of study. 
Namely, renters tend to avoid renting to people who are not of their own race.
The analysis also revealed that in all cities, the majority of hosts were Whites, even in cities with significantly more diverse racial composition. In some cities, Whites were over-represented compared to the local demographics, e.g., in Hong Kong 64\% of full property rentals hosts were whites even though 92\% of the population are Asians~\citep{koh2019offline}.

Additionally, according to \citet{cetre2020ethnic}, people have more trust in people of their own kind, i.e., those who are similar to them. 
In their study, \citet{cetre2020ethnic} examined ethnic in-group bias in United States and Germany through the Trustlab platform and showed that ethnic groups trust people from their own ethnic group more than those of other groups. They also showed that discrimination by the ethnic majority may be reduced when disclosing information on the trustee being rich, as it is removing the stereotype that ethnic minorities are ``undeserving poor.''

The film industry in particular has also been having a diversity problem \citep{Hollywoo99:online} evident in multiple aspects. \citet{smith2014race} showed that in 600 popular films (between 2007 - 2013), only about a quarter (25.9\%) of the speaking characters were of minority ethnic groups. Those characters are representatives of ethnic groups that comprise 37\% of the U.S. population and purchase 46\% of movie tickets, thus are obviously under represented. Moreover, when looking at the next generation to come, nearly 50\% of children under the age of five in the U.S. are not White, meaning that both the current and future audience for films is far more diverse than what is shown on screen. 
  \citet{smith2014race} also illustrated that existing cultural stereotypes may still govern how characters from different backgrounds are shown on screen. For example, Hispanic females seem to be displayed as more hypersexualized than their female counterparts from other ethnic groups.

  \citet{hennekam2018institutional} interviewed different workers of the film industry descrining how they quickly realized that being ``different'' from mainstream is an obstacle for their career advancement. Non-White women, as an example, appear unwelcome and less likely to succeed due to the absence of role models in the form of successful ethnic minority women in the industry~\citep{buunk2007positive}.

  According to the common notion among directors in the industry, casting minorities in lead roles is considered a financial risk~\citep{lee2014race}. 
Therefore, the ethnic background of the director may impact the actual casting - \textit{``Black directors conversely cast Black characters''}~\citep{smith2014race}.
Director Ridley Scott, who casted White actors for the top leading roles in a story featuring Egyptian characters, replied about his casting:
\textit{
``I can’t mount a film of this budget, where I have to rely on tax rebates in Spain, and say that my lead actor is Mohammad so-and-so from such-and-such […] I’m just not going to get it financed. So the question doesn’t even come up.''~\citep{lee2014race}.
}
\citet{lee2014race} studied this notion by investigating how a movie’s racial composition affects its performance. The results indicate no significant differences in performance between films that star White versus non-White leading actors. Hence, with the financial excuse out of the picture, it appears that the phenomenon of casting out minorities from lead roles in top movies, results from the lack in diversity awareness.

For a deep and broad study of the topic of ethnic diversity in the film industry, we developed and applied data science tools, including machine learning and deep learning. For those methods to work efficiently and accurately there is a need of large datasets with individuals labelled by ethnic origin.
According to a recent study, most existing large-scale face databases are biased towards “lighter skin” faces (around 80\%), e.g., White, compared to “darker” faces, e.g., Black \citep{merler2019diversity}.
Biased training data more likely to produce biased models that are trained on it \citep{mehrabi2021survey}. This in turn raises ethical concerns about the fairness of those models, which may eventually affect decision makers \citep{fletcher2021addressing}.

\subsection{Various Studies Relying on Poster Analysis}
Movie Posters may be considered as visual media that contain various explicit and implicit data on movies, like main actors and characters.
In the past couple of years, movie posters have therefore been of interest and analyzed for various studies.

\citet{aley2020powerful} conducted manual inspection on the main characters in movie posters of 152 popular U.S. animated childrens' films produced over the last 80 years. Their findings revealed that main characters were more likely to be male and that males were portrayed as more powerful.
\citet{gabriel2012rugged} also conducted a human-based examination of nearly 150 posters, all of which were on the yearly top-30 movies, as measured by gross box office profit (2007-2011). One of their key findings was that male characters outnumbered female characters by a ratio of three to one. This was accompanied by a bias toward male characters in the fundamental composition of the posters.

Other studies focused on specific ethnic groups through related posters. For example, \citet{freire2019cultural} investigated the current representation of the Latin American identity in mainstream media cinematic posters. \citet{freire2019cultural} utilised Latin American visual design language to reinterpret the possibilities that film posters have in creating elaborate narratives that treat audiences with respect and complexity.

Various studies performed manual analysis of posters of a specific movie or franchise, for instance:
\begin{itemize}
  \item \citet{rahmasari2014semiotic} analysed the posters of "The Help" movie that tells a story about racism.
  \item \citet{maiorani2007reloading} analyzed ``The Matrix" trilogy promotional campaign (1999–2003), focusing on the interplay between verbal and visual semiotics.
  \item \citet{nuraisyah2016semiotic} observed implicit meaning in 16 posters of the Harry Potter franchise, by inspecting different elements in the posters such as semiotic triadic, tagline and how the posters represent fantasy movies.
\end{itemize}

As another example, \citet{wi2020poster} used movie posters to assist in the classification of movies into genres using convolutional neural network (CNN). They extracted style features from the posters and identified correlations between the genres. That enabled them to obtain more accurate multi-genre classification. Their dataset consists of 20,764 poster images which were crawled in the order of box office and labeled by genre using IMDb.

To the best of our knowledge, a multi-ethnic analysis of posters was not yet conducted. Therefore, our study is a pioneer in performing a multi-feature multi-ethnic large-scale analysis of movie posters.

\section{Methods and Experiments}
\label{sec:method}

Our main goal of the study is to examine how different ethnic populations are represented in the film industry and whether there are objective indications of inequality and under-representation. We analyzed movie poster images from multiple genres to explore the current state of ethnic diversity and to test whether diversity awareness has changed over time.

\subsection{Datasets} \label{subsec:data}

We curated a large dataset, consisting of 148,748 posters images, used for all of the analyses further described in this paper. We collected the images and metadata from online publicly available sources, by the process described below.

\subsubsection{Movies Dataset}
\label{sec:imdb-tmdb-ref}
To study bias as it presented on movie posters, we assembled large-scale datasets of movies using the following publicly available data sources:

\begin{itemize} 
 
     \item  \textbf{IMDb} \citep{IMDB} is an online database of information related to films, television programs, home videos, video games, and streaming content online – including cast, production crew and personal biographies, plot summaries, trivia, ratings, and fan and critical reviews. As of August 2021, IMDb has approximately 8 million titles (including episodes) and 11 million personalities in its database. It also contains over 580,000 movie records along with 1,636,604 posters images.
     
     We downloaded IMDb's open movies dataset \citep{IMDbDatasets}, selecting the following information:
\begin{itemize}
    \item General movies metadata such as primary title, genres, and release year.
    \item Movies rating metadata such as average ratings and vote count.
\end{itemize}
     
     We then filtered the dataset to contain only non-animated movies rated by at least 1,000 reviewers, thus obtaining a dataset that consists of 31,054 movies. 

    \item \textbf{TMDB} \citep{TMDB} is a community built movie and TV database, which additionally to movie's metadata offers high resolution posters and fanart. As of August 2021, over 1,000 images are added every single day on average. TMDb officially supports 39 languages along with extensive regional data and every single day TMDb is used in over 180 countries. As of August 2021, TMDb contains 683,671 movie records with 2,873,601 images overall. We used TMDb to obtain data about the movie country of origin and posters and actor images.
\end{itemize}

\subsubsection{Posters Dataset}
\label{subsec:posters-data}
For each movie in the IMDb database, we fetched all its posters from the TMDB database, and the main poster from IMDb using their API's \citep{IMDbPY,TMDBapi}. This process yielded a total of 286,654 posters. Manual inspection of random movies suggested that many posters of a particular movie were duplicates of the same image, except for the language of the text (title and actors names).
To overcome these duplicates, we applied the dhash algorithm~\citep{dhash} to identify visually similar images. For instance, many duplicate posters contained the same graphics with the title in a different location. We calculated each image dhash value and computed the Hamming distance \citep{hamming1950error} between all pairs of images of a specific movie. 
We removed all images with a distance smaller than 16 - an empirically set threshold.\footnote{We randomly selected pairs of posters from the same movie and measured the distance between each pair. We then visually inspected the pairs and chose a threshold set by the average distance between duplicate posters, and validated it on the pairs of non-duplicates.} This step resulted in the final dataset of 125,439 non-duplicate posters that we used for this study.

\subsubsection{Actor Dataset}
For each movie in the movies' dataset, and each actor in the cast list of the movie, we used IMDb API to fetched the actor's name, the identifier ``imdb id" and credits ranking (position in the cast list). Next, for each actor in the Actor dataset, we used IMDb and TMDB APIs to download up to three main profile pictures. We subsequently integrated all collected images with the collected related metadata into one unified Actors' dataset containing 118,136 actors and 217,575 related images of the actors. We filter all grayscale photos by calculating the mean squared error for each channel from the average pixel value.
We filter grayscale images since it is harder to utilize them to identify the actor's ethnicity.
After filtering grayscale images the actor dataset contained 101,873 actors and 179,858 actor images.

\subsubsection{US Demographics Dataset}
The U.S demographic data is collected as part of the US census \citep{CensusBureau} that takes place every 10 years.
In terms of ethnicity, five categories are considered: White, Black or African American, American Indian or Alaska Native, Asian, Native Hawaiian or Other Pacific Islander.
Since the census considers Indian people as Asian (as being part of continent Asia), we accordingly merged our collected data of Indian and Asian when using the census data.

\subsection{Feature Extraction by Image Processing of the Collected Posters}   \label{sec:features}

\begin{figure}[t]
    \centering
    \includegraphics[width=0.45\columnwidth]{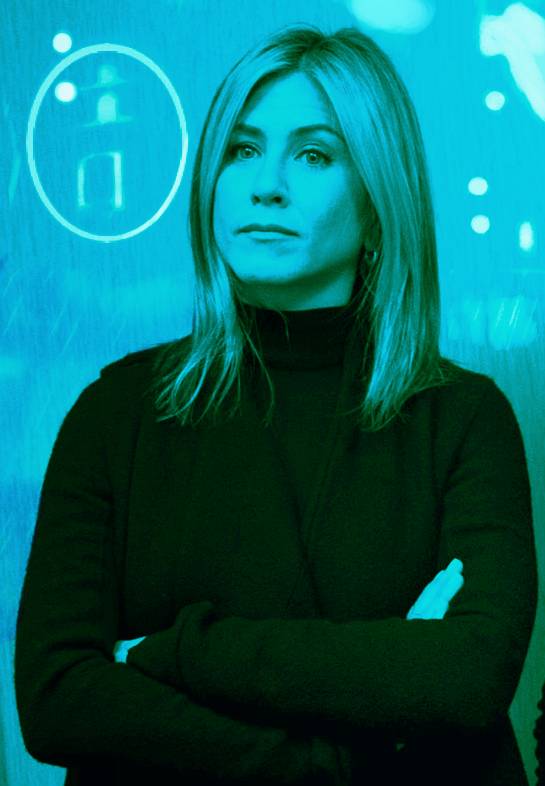}
    \includegraphics[width=0.49\columnwidth]{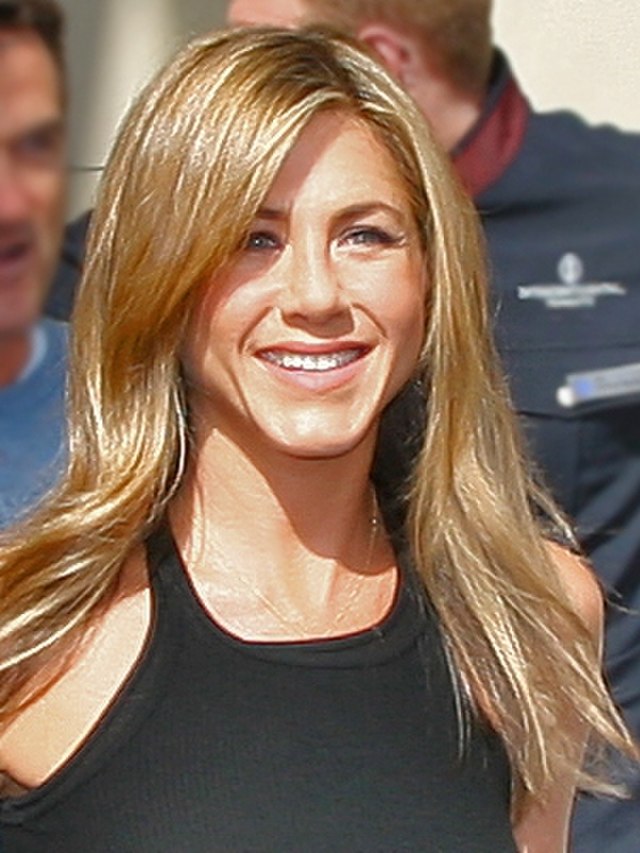}
    \caption{An example depicting the complexity of directly inferring ethnicity of actors from poster images. On the left: The actress Jennifer Aniston appears on a well-taken headshot profile photo. On the right: the same actress appears on a image where his face and skin color have been colorized.
    The image is based on the style of a real \href{https://www.themoviedb.org/t/p/original/tZR2n4PehYrMzva6QPdwOIDQ8x4.jpg}{poster} of the movie ``Blade Runner 2049" and is used in order to not violate copyrights. }
    \label{fig:poster-face}
  \end{figure}

In order to extract features representing actors in posters we used image processing of posters and profile images of the actors, in the following manner:
\begin{enumerate}
  \item \textbf{Posters Face Detection} - For each poster image, we applied the RetinaFace face detection algorithm \citep{deng2020retinaface}. We filtered out posters that did not contain at least one face, eventually holding a dataset of 77,192 posters images (about 60\% of the collected posters contained at least one face image). 
  
  \item \textbf{Actors Face Detection} - For each actor in the established actor dataset, we applied RetinaFace face detection algorithm \citep{deng2020retinaface} to identify the actor's face from the profile pictures. We then embedded each face using Arcface \citep{deng2019arcface}.
  Face embedding allows performing face verification by matching faces between multiple photos.
  
  \item \textbf{Actors Ethnic Recognition} - 
  For each poster, our goal was to determine the ethnicity of the actor faces depicted in the poster.
  However, using actor's faces from the poster is challenging and does not always yield a a good classification. This is because advertisers that create the posters usually use visual methods to engage potential viewers, methods which include unnatural backgrounds, mixed colors, and faces positioned in unclear angles (see Figure \ref{fig:poster-face}). 
  We, therefore, flipped the question - for each poster, we first recognized the participating actors and classified their ethnicity instead of ethnic-classifying the actual face we detected in the poster.
  
  Ethnic classification was performed using FairFace based models \citep{karkkainenfairface}.
  FairFace is a novel face image dataset with a balanced race composition for seven race groups: White, Black, Indian,\footnote{Indian refers to people who are originally from the Republic of India} East Asian, Southeast Asian, Middle East, and Latino. 
  This dataset enables better generalization and performance of classification for gender, race, and age, compared to model training (such as UTK~\citep{zhang2017age}, LFWA \citep{liu2015deep}, and CelebA \citep{liu2015deep}) on existing large-scale in-the-wild unbalanced datasets. 
  FairFace provides two models of ethnic classification, one of which classifies faces into four\footnote{Asian, Black, Indian, White} and the other into seven\footnote{White, Black, Latino-Hispanic, East Asian, Southeast Asian, Indian, Middle Eastern} racial groups.
  
  For each profile picture, we computed the actor's ethnic scores. The models provide an ethnic score of an actor per ethnic group (the probability of being part of the ethnic group). Since we collected up to three profile pictures per actor, we averaged the ethnic scores computed per each profile picture and then selected the max averaged score as the voted ethnic group.

  \item \textbf{Actors Recognition in Posters -} To match between posters and actors, we encoded each detected face (both from the actors headshots and the posters) by utilizing the ArcFace \citep{deng2019arcface}. Using ArcFace we encode all the faces detected in the posters and the actors photos.  Once we had all the faces encoded, we matched the encodings of the faces detected in the posters to the database of actor face encodings. We searched for the closest actor face by euclidean a distance. Then we evaluated two approaches for matching posters with actors (see Section \nameref{subsec:poster-actor})

\end{enumerate}

\subsection{Evaluation of the Models in Use}
\label{subsec:eval}
To evaluate the performance of the machine learning models used to generate the datasets for our study, we performed the following steps:

\begin{enumerate}

    \item \textbf{Face Detection Algorithm} - 
    To validate face detection, we manually inspected a random sample of posters containing a total of over 100 faces.

    \item \textbf{Face Recognition} \label{actor-recognition-algorithm} - To validate  performance of the face recognition algorithm, we randomly selected 50 posters to extract 149 faces. For every face, we created a new picture as a collage of 2 images: 
    \begin{enumerate}
        \item The original poster with a painted rectangle around a specific detected face.
        \item The headshot of the actor who was matched to the detected face.
    \end{enumerate}
    We then manually evaluated the matches.

    \item \textbf{Ethnic Classification Algorithm} - 
    To validate FairFace~\citep{karkkainenfairface} performance on the data used in this study, we selected random images of actors and actresses from each race category considered in our models. Using IMDb website, we verified their ethnic origin and tagged them accordingly.
     A total of 40 actors were selected for evaluating the four-race model and 70 actors for the seven-race model (5 actors and 5 actresses for each race category).
    We used the race predictions of the actors using the Ethnic Recognition algorithm (see Section \nameref{sec:features}) and manually evaluated the predictions.
\end{enumerate}

\section{Results}
\label{sec:results} 
To explore ethnic diversity from a new angle we analyzed 77,192 posters of 24,062 movies.
We searched for ethnic related trends in actor casting across decades. We also explored differences between English and Non-English foreign movies (see \nameref{sec:Appendix}). The results of our analysis are as follows.

First, we evaluated the ML models used in this study. The RetinaFace face detection algorithm correctly detected 100\% of the sampled faces.
To match posters to actors, we have evaluated two matching approaches (see Appendix, Section \nameref{subsec:poster-actor}), top-10 and all actors approaches. Both approaches for matching posters, yielded 100\% verification (every face was automatically matched to the correct actor, verified manually). The two methods differed slightly in their identification score (number of matched faces out of the total detected faces): 71\% for the whole cast list approach and 69\% for the top-10 approach got.
Evaluating ethnic classification methods (see Table \ref{table:1}), we found that the four race model provided an average precision of 92.85\% while the the seven race model only an average precision 61.37\%.

\begin{figure} 
         \centering
        \includegraphics[width=0.5\textwidth]{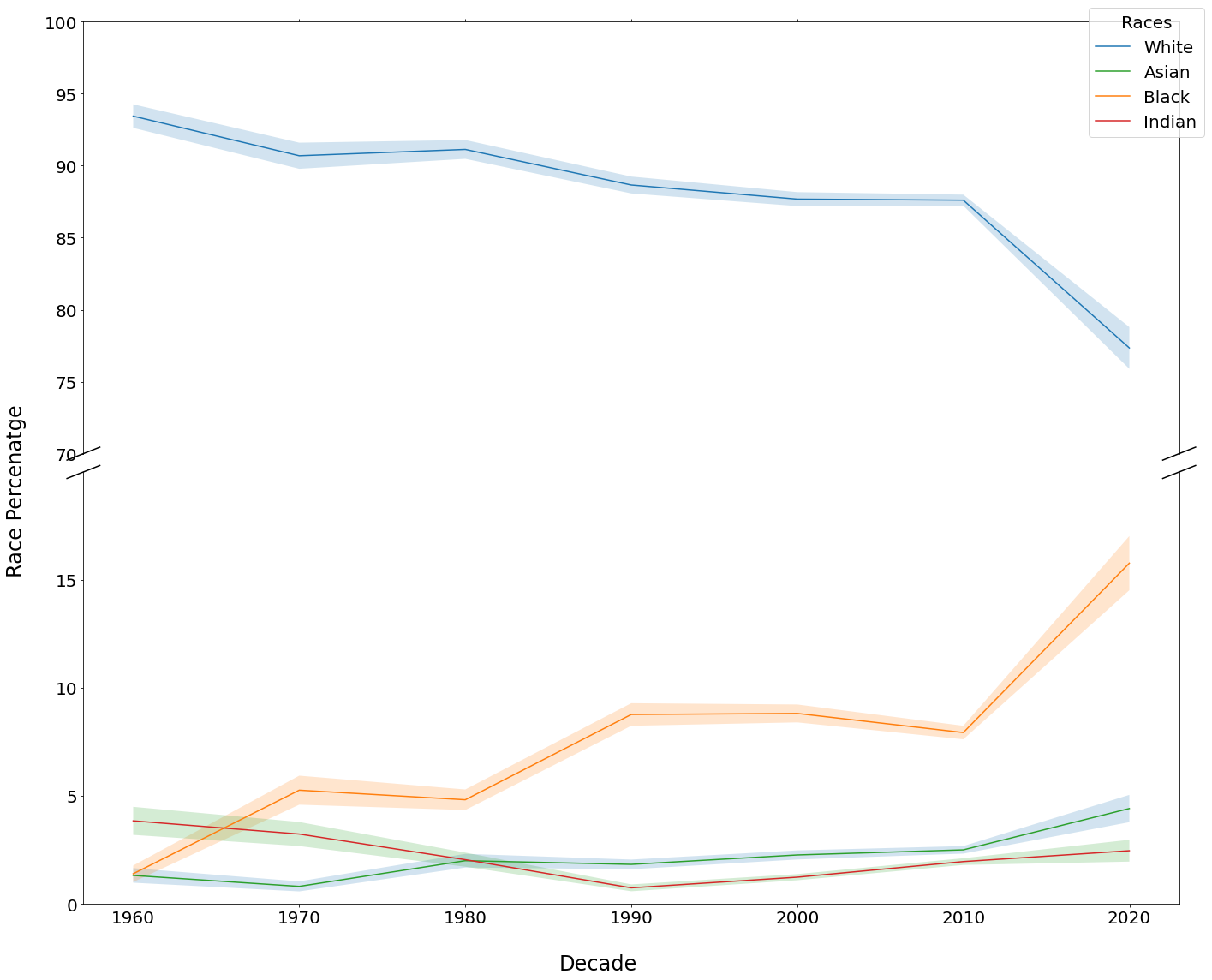}
        \caption{Trends in the relative frequency of ethnic actors representation in posters. The relative representation of actors from each ethnic groups is displayed as the percentage of ethnic face appearance compared to the overall number of faces detected.}
        \label{fig:num-face-races-year-decade}
\end{figure}

\begin{figure} 
     \begin{subfigure}[t]{0.49\columnwidth}
         \centering
        \includegraphics[width=\textwidth]{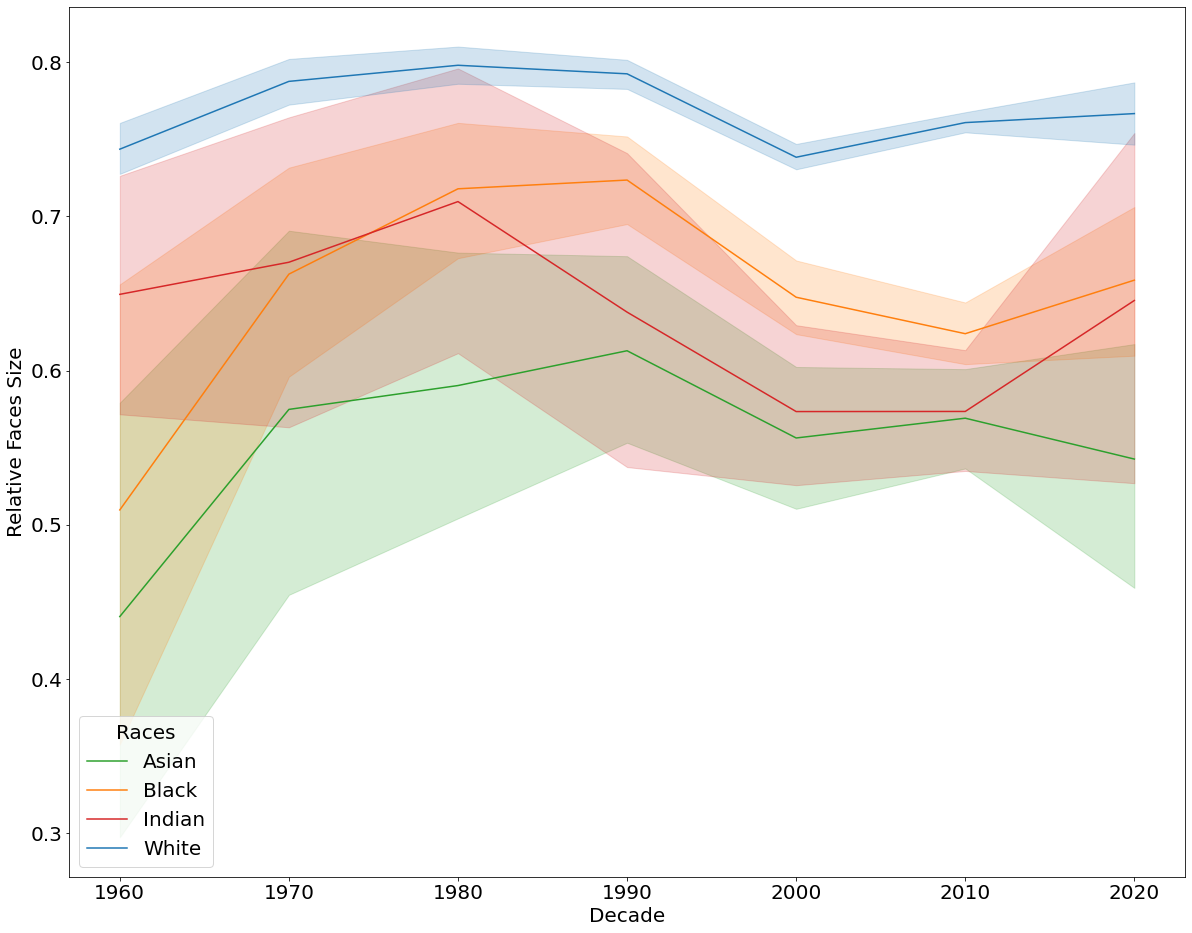}
        \caption{The change in the averaged face size relative to the largest face detected, for the different considered ethnic groups.}
        \label{fig:face-relative-year-decade}
     \end{subfigure}
     \hfill
     \centering
     \begin{subfigure}[t]{0.49\columnwidth}
         \centering
        \includegraphics[width=\textwidth]{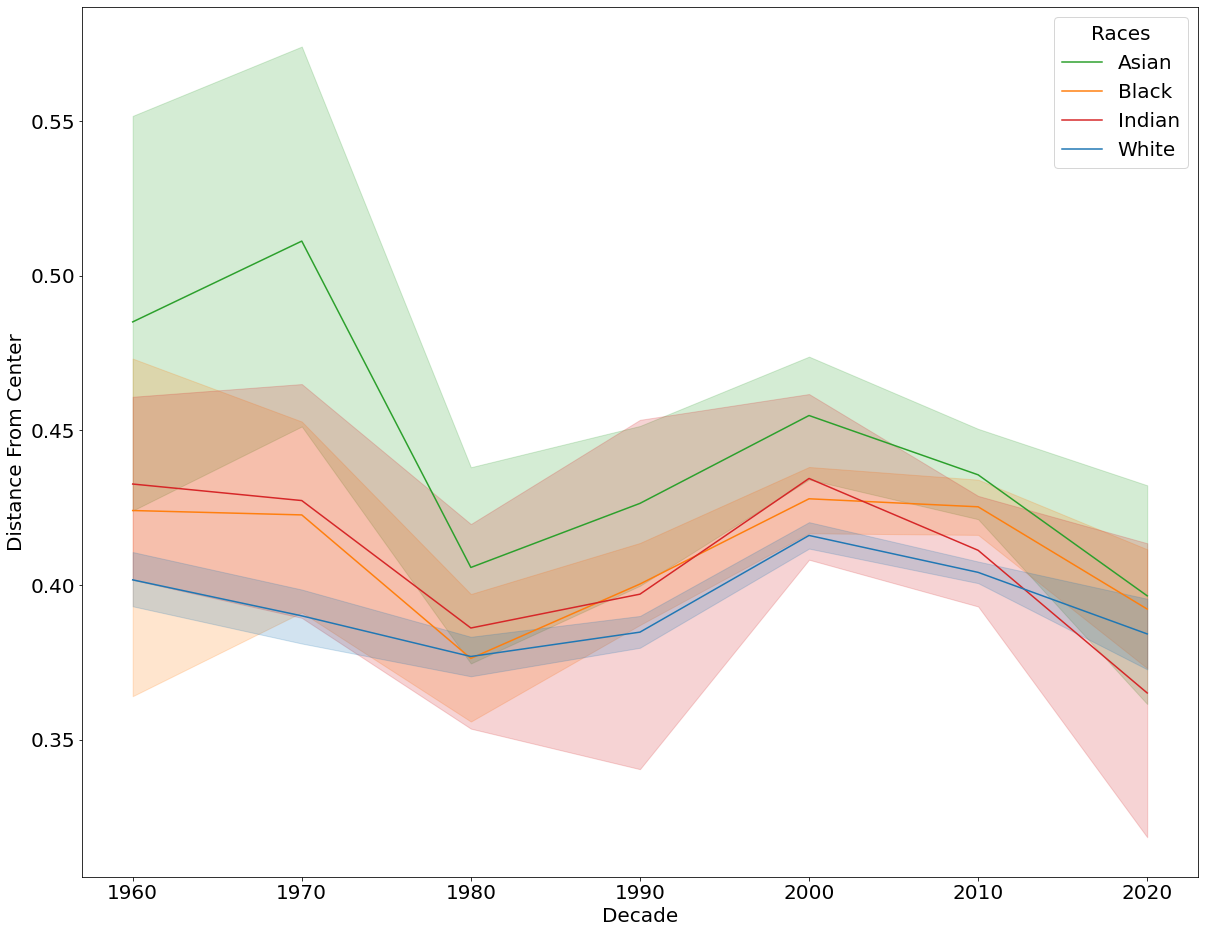}
        \caption{Normalized euclidean distance from the actor to the middle of the poster.}
         \label{fig:face-center}
     \end{subfigure}
  \caption{Trends in race appearance on posters.}
\end{figure}

Having tested the models, we proceeded towards exploring trends in race appearance on posters.\footnote{All the the calculation are made at a movie level to reduce noise that is generated by movies with high amount of posters.} 
To study ethnically related trends in the relative frequency of actor representation on posters, we analyzed the number of faces from each category and displayed it as the percentage of the overall number of faces in a poster (see Figure \ref{fig:num-face-races-year-decade}). 
We find that there is a clear White dominance. Even in the current decade, where the minimum value is seen, there are still 77\% White actors on movie posters. Nevertheless, there has been a constant increase in the relative representation of other ethnic groups over the years. 
In the last decade, the percentage of black actors almost doubled, and there was also significant growth in Asian actors representation.
Additionally, we normalized Figure ~\ref{fig:num-face-races-year-decade} data by demographic data per each ethnic group, according to USA ethnic population distribution,\footnote{There is no worldwide ethnic demographic data available.} as extracted from publicly available databases (see Section \nameref{subsec:data}) \citep{gibson2002historical, grieco2001overview, humes2011overview}.
We found that nowadays (2020-2021), there is almost a perfect balanced representation of minorities on posters according to their demographic distribution.

\begin{figure}
         \centering
        \includegraphics[width=\columnwidth]{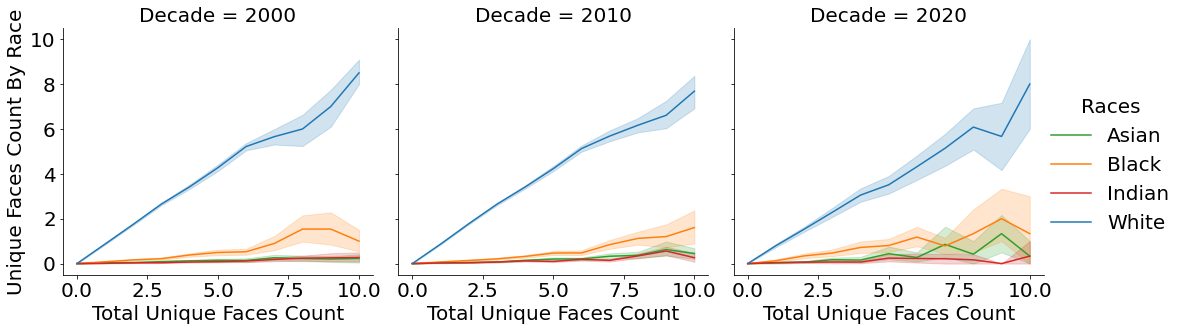}
        \caption{The average number of unique actors on English movie posters 2010-2021.}
         \label{fig:num-faces-decade}
\end{figure}

We then studied the ethnic effects of actor face size and location on the posters.
We specifically evaluated how the size of actor faces relate to the size of the face of the largest actors on the poster (see Figure ~\ref{fig:face-relative-year-decade}).
The incentive for exploring the representation of face sizes is that larger faces should represent more important actors or characters.
We found that the average relative face size of white actors 25\% larger than the other races. 
Also, we see similar trends in terms of distance from the poster center (see Figure \ref{fig:face-center}).
We also observed that in the past couple the difference has reduced drastically.

Next, we inspected the relationship between the number of different actors on movie posters and ethnicity (see Figure \ref{fig:num-faces-decade}). 
We have observed a growth in minority representation on the posters in the past 22 years. The growth is most evident for black actors.
However, we see that minorities have higher representation, primarily in movies with more than six different actors on the movie posters. 
In other words, minorities have a higher chance of appearing on a poster of movie where there many actors appearing on the movie posters.

 \begin{figure} 
         \centering
        \includegraphics[width=\columnwidth]{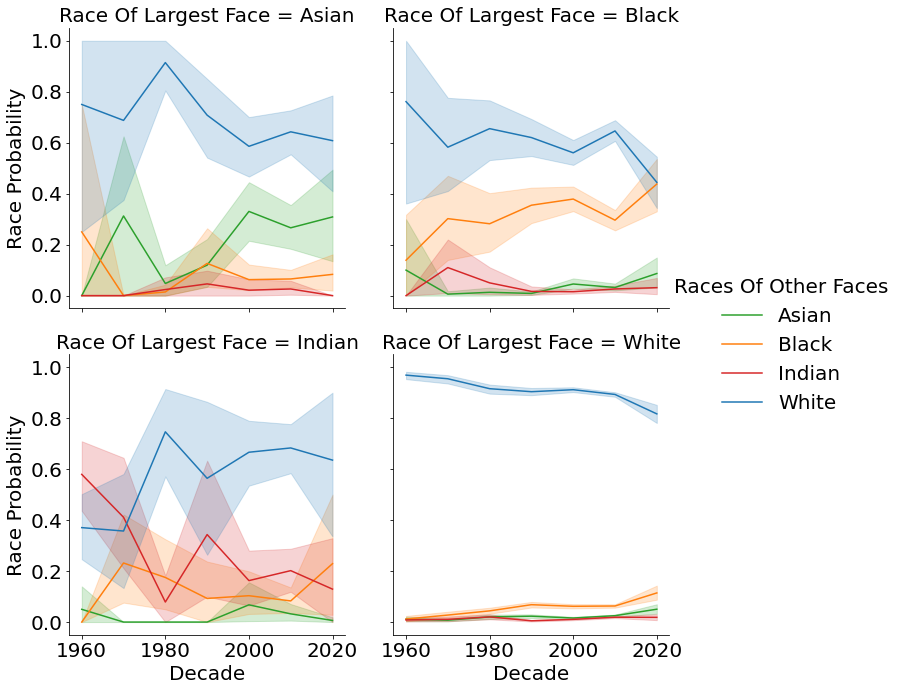}
    \caption{Changes in the conditional probability of a face in a poster to belong to each race category, given the race of the largest face in the poster.}
    \label{fig:races-dist-given-largest}
\end{figure} 

Additionally, we explored if the selection of the second largest character on a poster is race biased, conditioned by the ethnic origin of the largest character on the poster.
Thus, for each face, we analyzed its probability of belonging to each of the ethnic categories, given the race of the largest depicted face in the poster (see Figure \ref{fig:races-dist-given-largest}).
We found that White actors have the highest probability of being second across all four graphs, meaning that no matter the ethnicity of the largest actor, the rest of the actors are most likely White. 
Moreover, when the largest actor (largest face size in the poster) is of Non-White, the next most probable race category for actors (after White actors) is the same as the largest face on the poster.

\begin{figure*}
     \begin{subfigure}[t]{0.49\textwidth}
         \centering
        \includegraphics[width=0.95\textwidth]{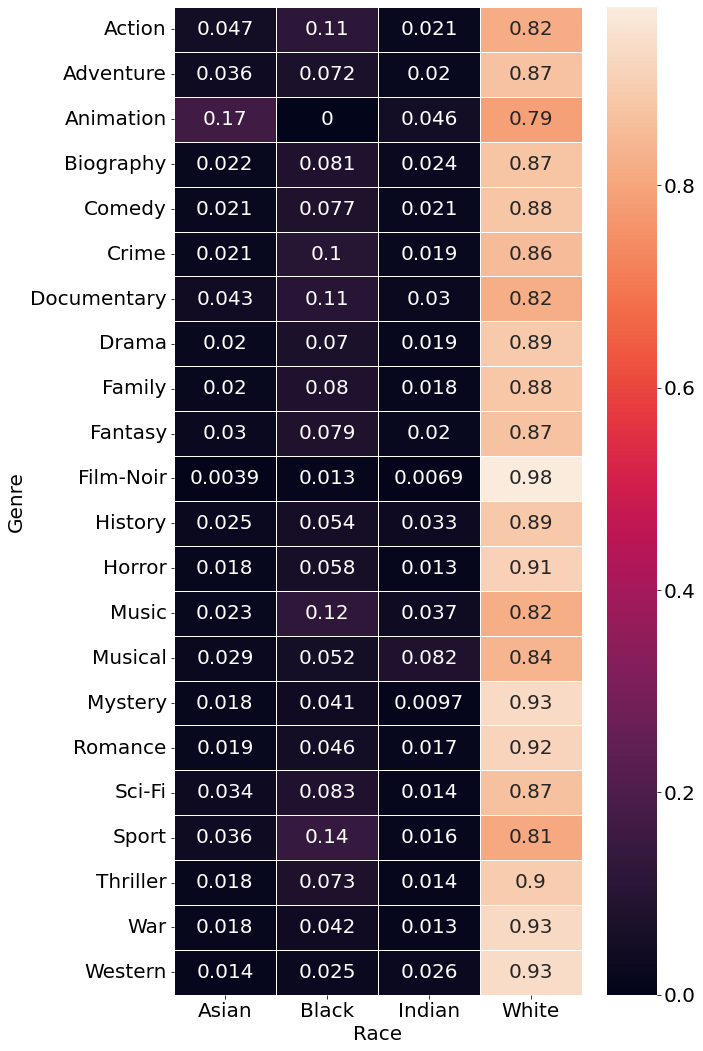}
    \caption{Racial distribution for each genre.}
    \label{fig:heatmap-genres-faces-races-count-all-years}     \end{subfigure}
     \hfill
     \begin{subfigure}[t]{0.49\textwidth}
         \centering
        \includegraphics[width=0.95\textwidth]{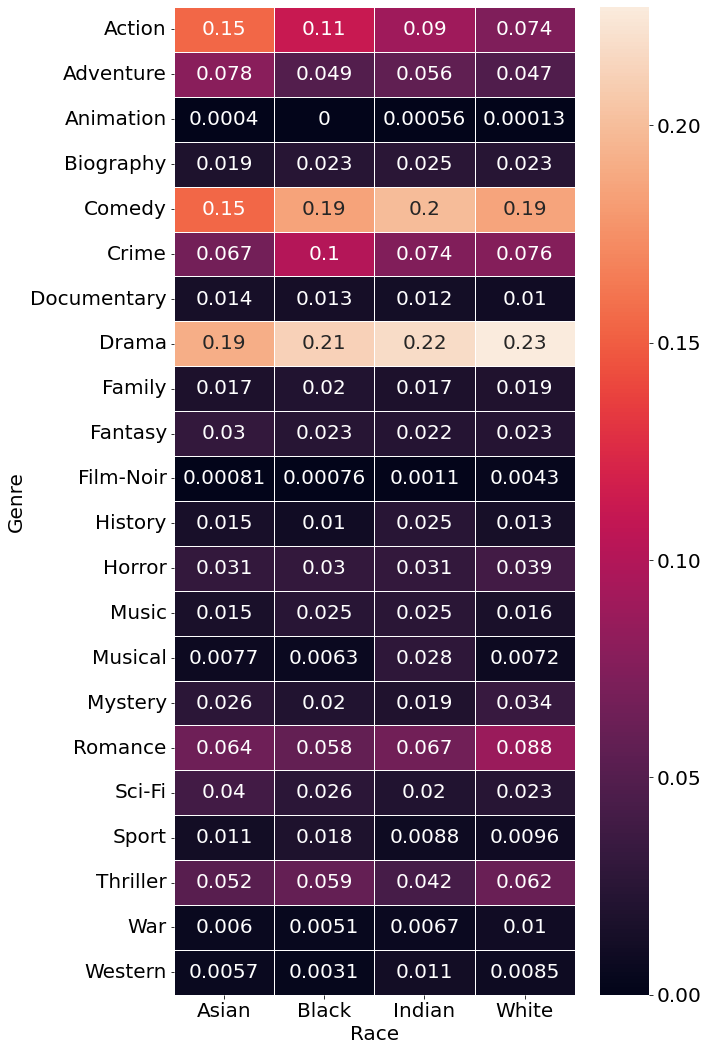}
    \caption{Distribution for each race of appearance on posters by genre. }
    \label{fig:race-genre-dist}
     \end{subfigure}
     \hfill
     \centering
     \caption{Heatmaps for probability of ethnic actors appearance in movie posters.}
\end{figure*}

We further analyzed the data for different movie genres. 
For each genre available in the IMDb dataset, we inspected the ratio of racial distribution in its posters (see Figure \ref{fig:heatmap-genres-faces-races-count-all-years}).
White actors have the highest percent appearance across all genres and are particularly dominant in the Film-Noir category, holding a maximum value of 0.98 and in the Western and Mystery genres (with 94\% and 93\% respectively).
The Sport, Music, Action, Crime, and Documentary genres include the highest percentage of Black actors, and the other two minority ethnic categories (Asian, Indian) have meager representation, with a maximum value of around 9\%.
Additionally, we inspected for each race its distribution across genres (see Figure \ref{fig:race-genre-dist}).
We observe that Asian actors are more apparent in the action genre and Black act than other races.

\begin{figure} 
         \centering
        \includegraphics[width=\columnwidth]{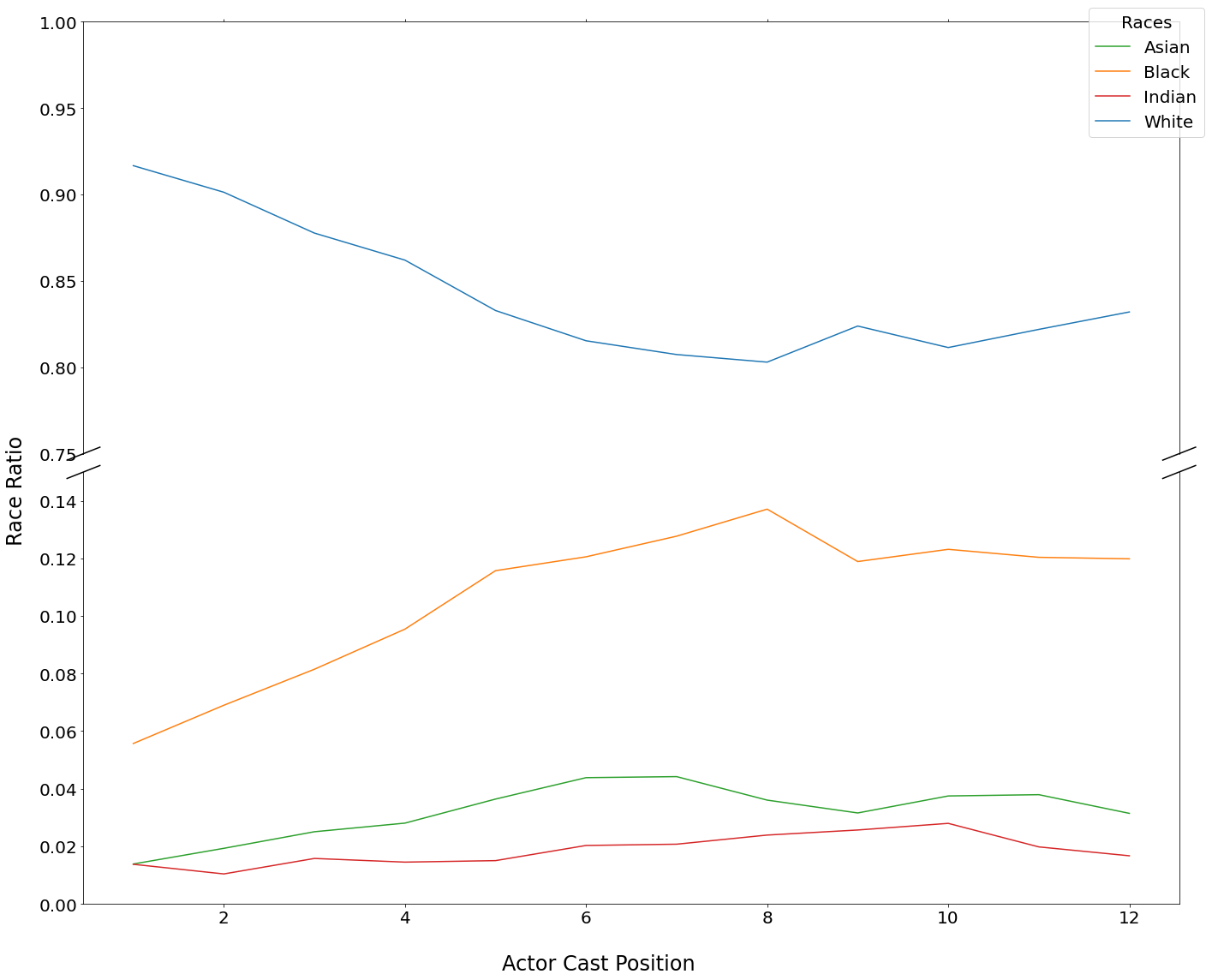}
    \caption{Ethnic Related Appearance in each Rank Position: at each rank, the relative number of actors of each race category out of all actors is depicted.}
    \label{fig:actor-rank-race-ratio}
\end{figure}

Finally, we examined the first 12 rank positions in the cast list. We define rank as the sequential number within the cast list.
We counted the number of actors and divided them into the different race categories (Asian, Black, Indian, and White). For each rank, we calculated the ratio between the number of actors of each race with respect to the total number of actors at that rank (see Figure \ref{fig:actor-rank-race-ratio}).
The results demonstrate that White actors are the most common throughout ranks in posters. Moreover, in English posters, the White actors relative appearance declines gradually with rank. In contrast, Black actors relative appearance gradually increases with rank, indicating White actors are preferably positioned high up within the cast list, unlike Black actors who are positioned in more minor roles. Other ethnic groups demonstrate consistently low presence across all rank positions in the cast list.

\section{Discussion}
\label{sec:dis}
We developed an empirical analysis framework to study ethnic bias in the movie industry through the eyes of a poster viewer.
To this end, we collected a large-scale dataset of movie posters from the last 60 years.
Using deep learning algorithms, we extracted features from the posters and explored the racial bias in the film industry from multiple perspectives.

We studied trends in the ethnic-related representation of 26,069 actors in movie posters. We found that the representation of White actors in English movies tends to decrease with time, in parallel to an increase in representation of minorities.
These results are strengthened by a report on evidence of steady growth in minority representation in movies~\citep{HollywoodDiversityNewsroom}.
We speculate that these changes result from rising cultural awareness. The BLM movement that started in 2013 can explain the drastic increase in black actors' representation in the past decade.
Additionally, Hollywood has targeted the global market in the past decade, especially China \citep{Hollywoo95:online}.
As expected, we have seen a drastic growth in Asian actors' representation in the past decade.
However, we see a different picture from the same data when it normalized relative to the U.S. population. 
We observe a similar increase in representation for all inspected ethnic minorities.
Surprisingly, today, poster representation achieved equilibrium relatively to their population in the U.S, and there is no over-representation to appeal to specific markets.
These results suggest that today, the film industry makes an active attempt to put together posters that fairly represent the U.S. population according to ethnic distribution.

We hypothesized that minorities' size and position on posters would depend on their ethnicity. 
We found that white actors are larger and closer to the center of the poster.
This might be explained by the fact that white actors still get more lead roles since they are the biggest ethnic group of U.S. actors. 
Leading roles usually translate to larger characters on posters and more central positioning.

Also, we were interested in testing whether in recent years actors of color were added to posters from more minor roles to appeal to minorities.
As we observed on movie posters from the last couple of decades, minorities are prominently featured on movie posters with many different actors. 
We suspect that there are several possible reasons. The first option is that movies franchises with many stars such as Marvel, DC, Fast \& Furious have become more common in the past two decades. 
Since these movies have many famous actors, many actors get placed on posters, including actors of color.
Another option is that movie production companies feel that adding minorities to movie posters with many actors will less critique.

Furthermore, we found evidence of homophyly on posters where the largest actor from a specific ethnicity there is a higher likelihood to find actors from the same ethnicity than on other posters. We suspect that many actors of color are cast to movies with an ethnically related plot, requiring many actors from the same ethnicity.

Also, we found that the most diverse genre in English movies is Documentary, where the highest percentage of minorities on posters was observed. 
This is likely because documentaries are true stories, which leave little control over the cast to the producers.
We did see indications for stereotypical representation of Asian and Black actors.
Relative to other ethnicities, Black actors are more dominant in crime movies while Asian actors are in action movies.
We suspect it is a result of the stereotypes that all Asians know martial arts \citep{AllAsian63:online, Asiansha23:online} and Afro Americans are criminals \citep{welch2007black}.

Finally, we found that in general, the number of actors appearing on posters distributes nearly exponentially with rank.
The higher the actor's rank in the cast list, the higher their probability of appearing on the poster. This result indicates that actor appearance on posters has an excellent potential to act as a centrality measure of leading characters. 
In terms of ethnic related effects on poster appearance at different ranks, we found that
the percentage of non-White actors on posters is higher on lower-ranked positions on the cast list. With the data at had, we are not able to conclude if it results from affirmative action or from the fact that non-White actors hold ranking in the cast list that is lower than their actual importance in the first place.
Diving into temporal trends in actor appearance on posters by cast list position and race, we find an overall increase in representation of non-White actors in most cases regardless of cast position in the past ten years.

As with any machine learning-based study, this work is mainly limited by the models used.
In this study, we used the available state-of-the-art algorithms to reduce errors.
Obviously, when better race prediction models will become available, also more accurate results could be obtained in the future.
We found this to be an issue, mostly in race prediction on older black and white movies, where the sample size of non-White faces were small.
The resolution of a poster is also a limitation, especially when there is a small face. A possible solution might be the use of super-resolution, but a further study should be performed to determine its effectiveness. 
Another limitation is that no data specify how many posters were produced for each movie and how many posters were printed from each variation which may portray a different picture.

\section{Conclusions}
Movie posters have an enormous potential for highlighting cultural biases, and in particular ethnic biases that appear in the film industry, and that eventually also shape our cultural perception as viewers. 
In this work, we created the first dataset of movie posters that contains metadata about actors' identities and ethnicity.
We constructed this dataset using deep learning models and fusing the poster data with multiple sources. We then analyzed the diversity in the film industry by examining multiple parameters.

Our results suggest that when it comes to poster design, on average white actors are larger and closer to the center of the poster. 
We also found an increase in the representation of Asian and Black actors in the past decade on posters.
In fact, in English films, the actors on recent posters from the past two years represent exactly the ethnic distribution of the U.S.
Additionally, we demonstrate that the main character's race affects the other characters' race on the poster, with greater probability for their own kind.
In a future study, we plan to compare posters from different countries, looking for differences between the presented posters in different cultures. Additionally, we intend to present a centrality measure for the rank in the cast list, based on movie poster appearance. In fact, it would be interesting to find cases of mismatch, where certain characters of minor roles are of greater appearance in posters, this might indicate an unfair attempt for correction of ethnic representation gaps.
\label{sec:con}

\section{Data availability}
All code and data will be available upon publication.

\section{Acknowledgements}
We thank Valfredo Macedo Veiga Junior (Valf) for designing the infographic illustration.

\bibliographystyle{agsm}
\bibliography{sample-bibliography}

\clearpage
\appendix
\section{Appendix}
\label{sec:Appendix}

\subsection{Poster Actor Matching} \label{subsec:poster-actor}
We evaluated two approaches for matching face to an actor: 
\begin{enumerate}
    \item Compare each face found in the poster to the whole cast list of the movie.
    \item Compare each face found in the poster to only the top-10 actors from the movie's cast list.
\end{enumerate}
Our line of thought for the second approach was that actors who appeared on the posters are most likely amongst the top-10 of the cast list; therefore, we believed the second approach would be more effective. 
We quantified the number of faces we detected across the posters and found that, on average, there are 3.81 actors on a poster with a standard deviation of 4.18. This indicates that most posters show actors from the top-10 most central roles. Hence, comparing faces to the complete cast list may add noise and cause matching errors. We validated this hypothesis (see Section \nameref{subsec:eval}) and showed that it was incorrect - the approach of the whole cast list was better in terms of accuracy. We noticed that due to damaged data in IMDb's datasets, the order of the cast list in some movies was arbitrary and main actors who appeared in the posters were not among the first 10 actors. Moreover, Arcface detection was so accurate that trying to match actors from the bottom of the list did not add noise as we first suspected. 

\subsection{Figures and Tables}

    \begin{table}[H]
        \centering
    \begin{tabular}{| c | c | c|}
     \hline
    Race & Precision & Recall  \\      \hline
        Asian & 100.0\% & 100.0\% \\
        Black & 100.0\% & 90.0\%  \\
        Indian & 100.0\% & 70.0\% \\
        White & 71.42\% & 100.0\% \\
         \hline

    \end{tabular}
    \caption{4-classes model evaluation.}
    \label{table:1}
    \end{table}
    
    \begin{table}[H]
        \centering
    \begin{tabular}{| c| c| c|}
     \hline
      Race & Precision & Recall \\      \hline

        Black & 100.0\% & 70.0\% \\
        East Asian & 60.0\% & 90.0\% \\
        Southeast Asian & 66.66\% & 20.0\% \\
        White & 55.55\% & 100.0\% \\
        Indian & 66.66\% & 40.0\% \\
        Latino Hispanic & 30.76\% & 40.0\% \\
        Middle Eastern & 50.0\% & 40.0\% \\
         \hline

    \end{tabular}
        \caption{7-classes model evaluation.}
        \label{table:2}
        \end{table}

\subsubsection{Foreign Movies}

\begin{figure} 
     \begin{subfigure}[t]{0.29\columnwidth}
         \centering
        \includegraphics[width=\textwidth]{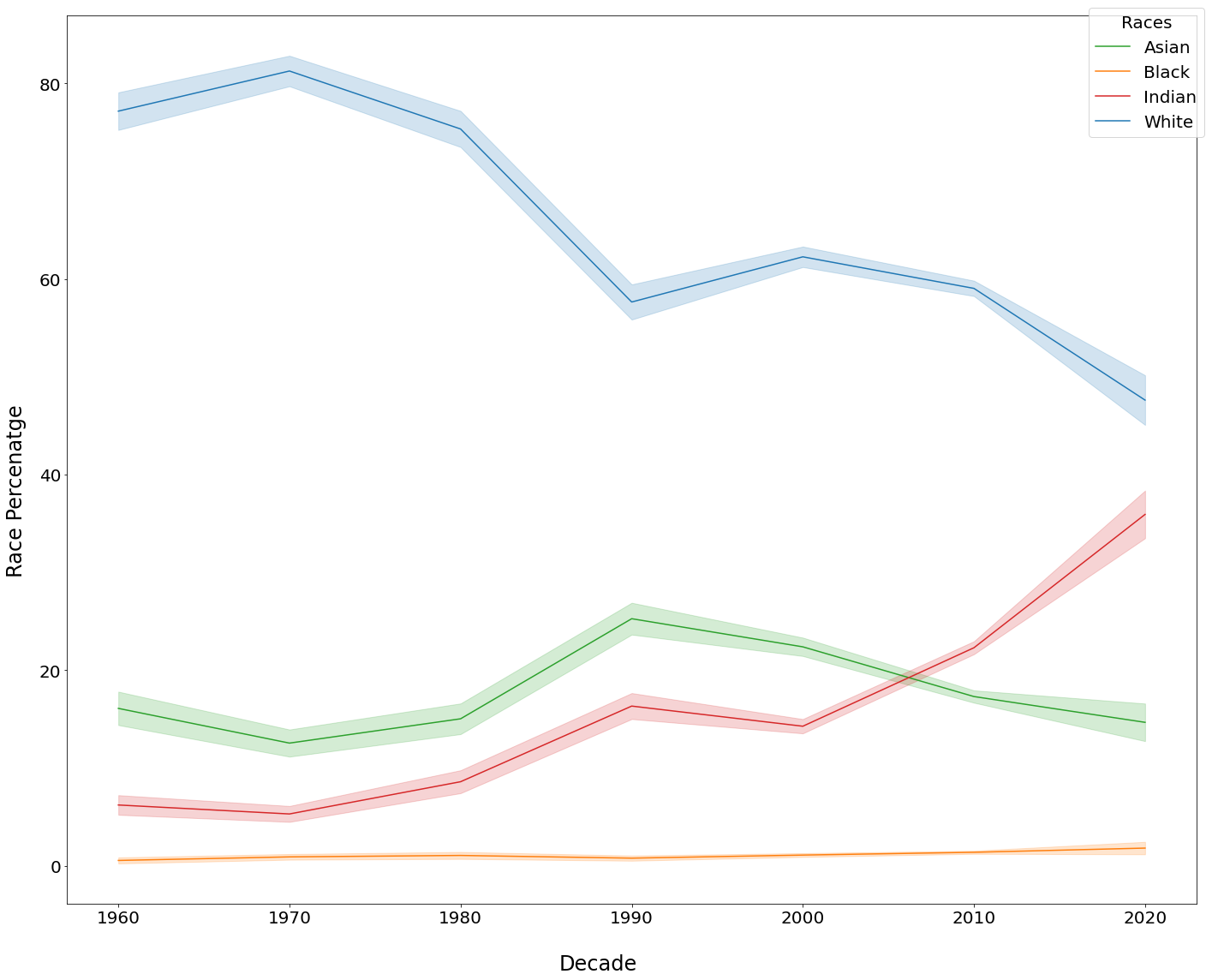}
        \caption{Trends in the relative frequency of ethnic actors representation in posters. The relative representation of actors from each ethnic groups is displayed as the percentage of ethnic face appearance compared to the overall number of faces detected.}
     \end{subfigure}
     \hfill
     \begin{subfigure}[t]{0.29\columnwidth}
         \centering
        \includegraphics[width=\textwidth]{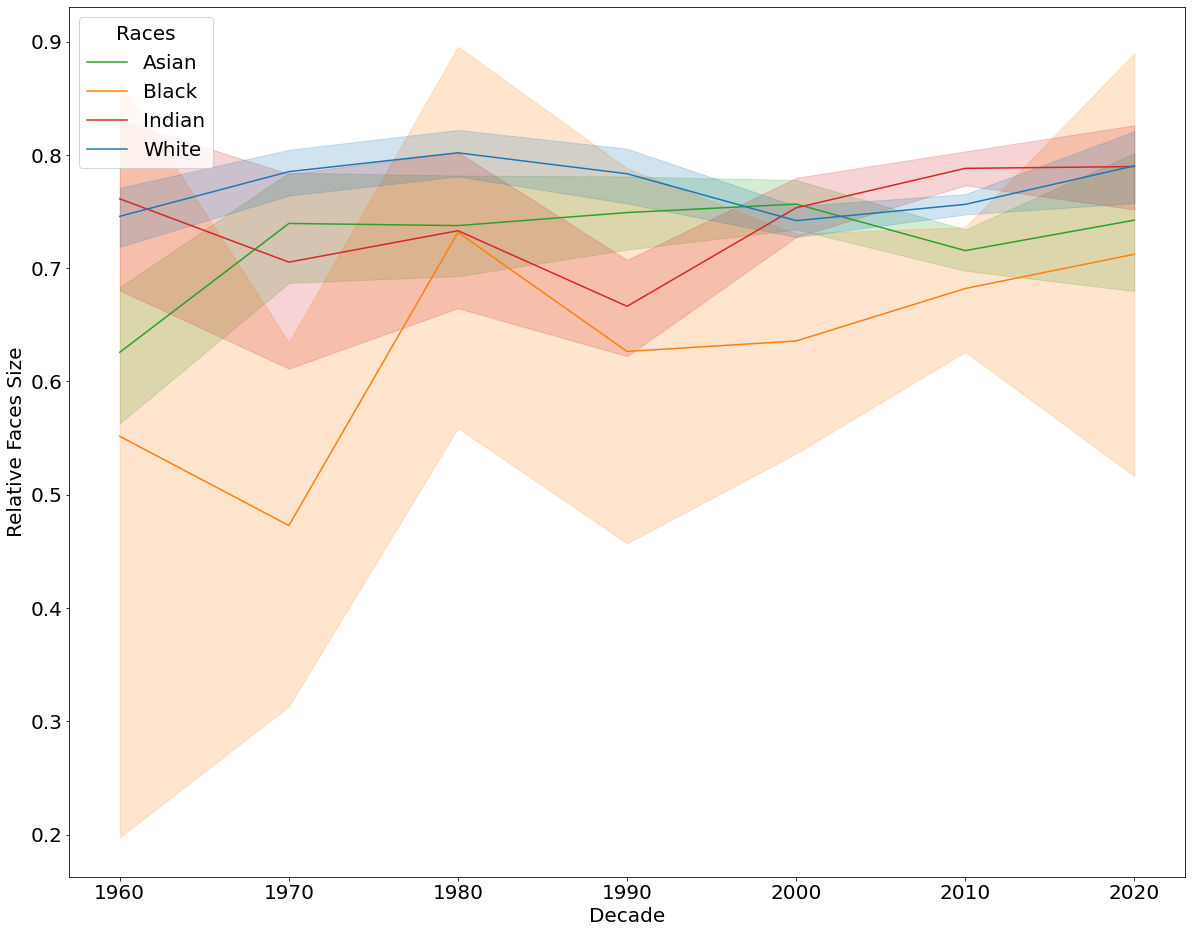}
        \caption{The change in the averaged face size relative to the largest face detected, for the different considered ethnic groups.}
     \end{subfigure}
     \hfill
     \centering
     \begin{subfigure}[t]{0.29\columnwidth}
         \centering
        \includegraphics[width=\textwidth]{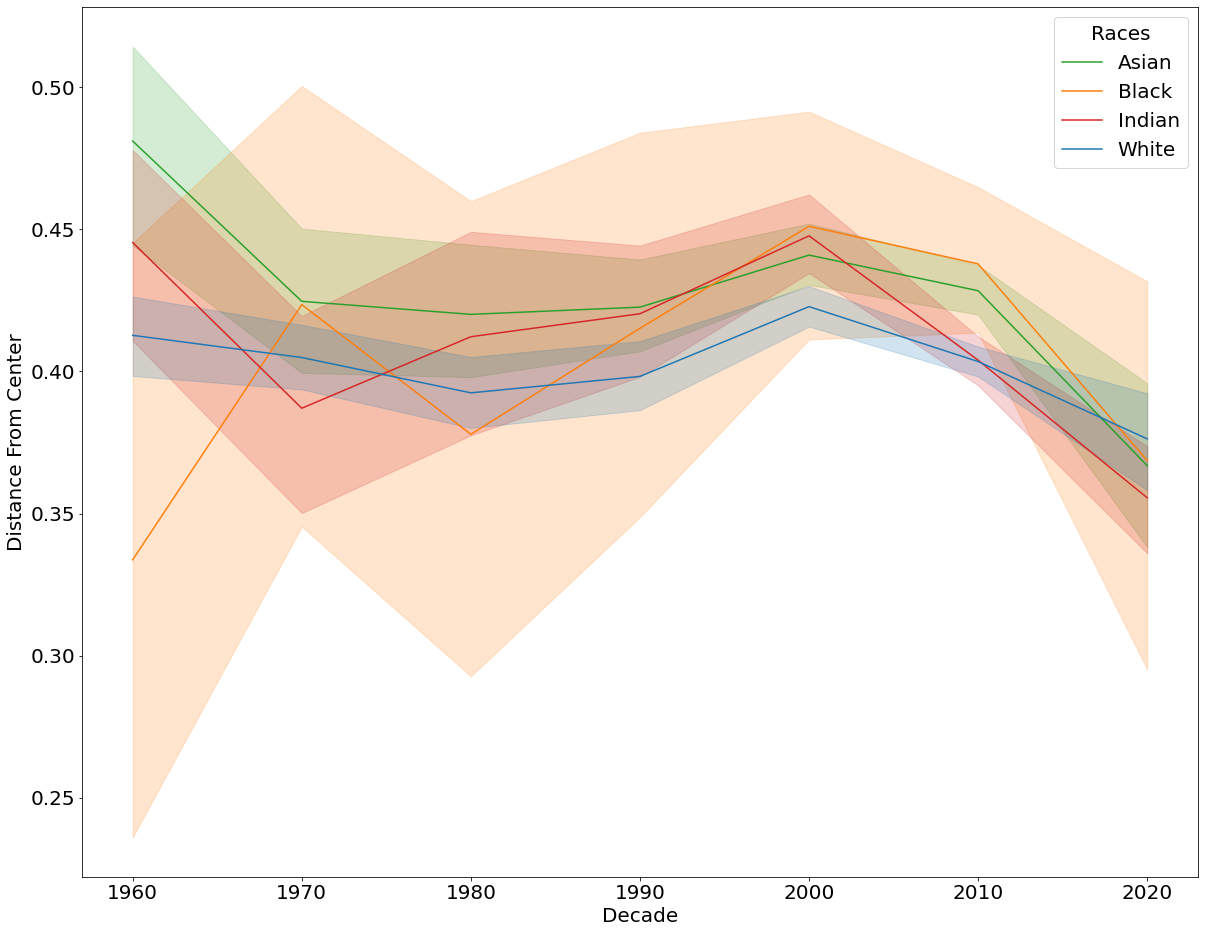}
        \caption{Normalized euclidean distance from the actor to the middle of the poster.}
     \end{subfigure}
  \caption{Trends in race appearance on posters.}
\end{figure}

\begin{figure} 
         \centering
        \includegraphics[width=\columnwidth]{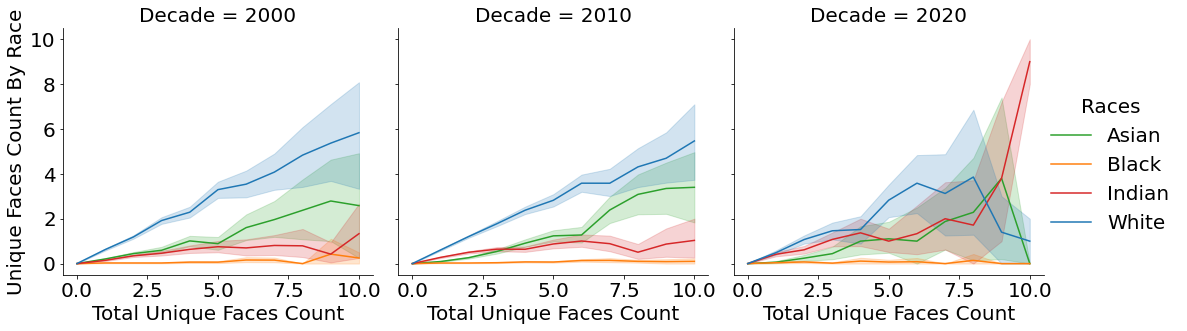}
        \caption{The average number of unique actors on English movie posters 2010-2021.}
\end{figure}

\begin{figure} 
         \centering
        \includegraphics[width=\columnwidth]{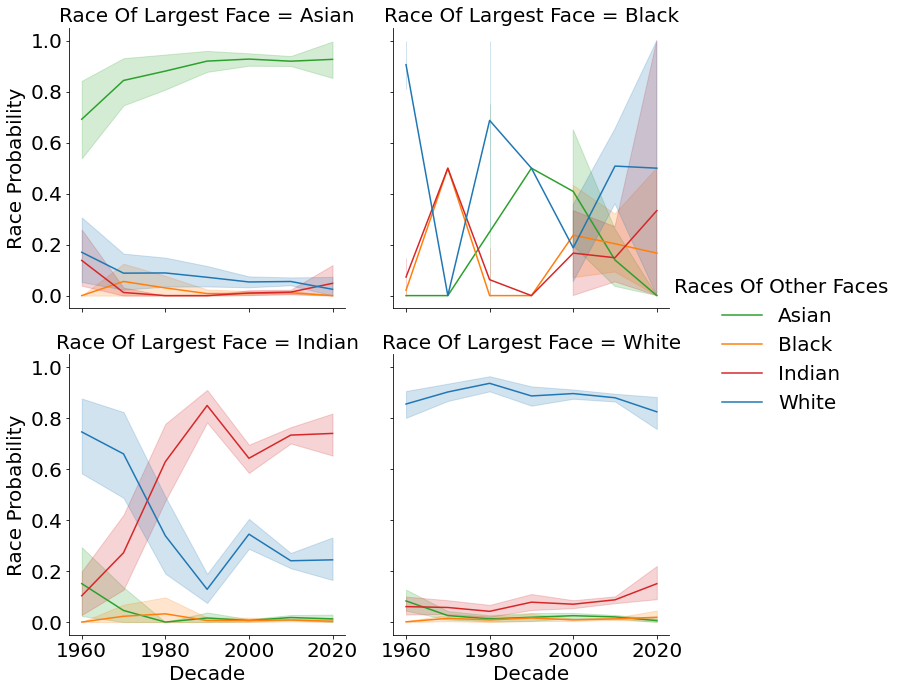}
     \hfill
     \centering
    \caption{Changes in the conditional probability of a face in a poster to belong to each race category, given the race of the largest face in the poster.}
\end{figure}

\begin{figure*}[t]
     \begin{subfigure}[t]{0.49\textwidth}
         \centering
        \includegraphics[width=0.95\textwidth]{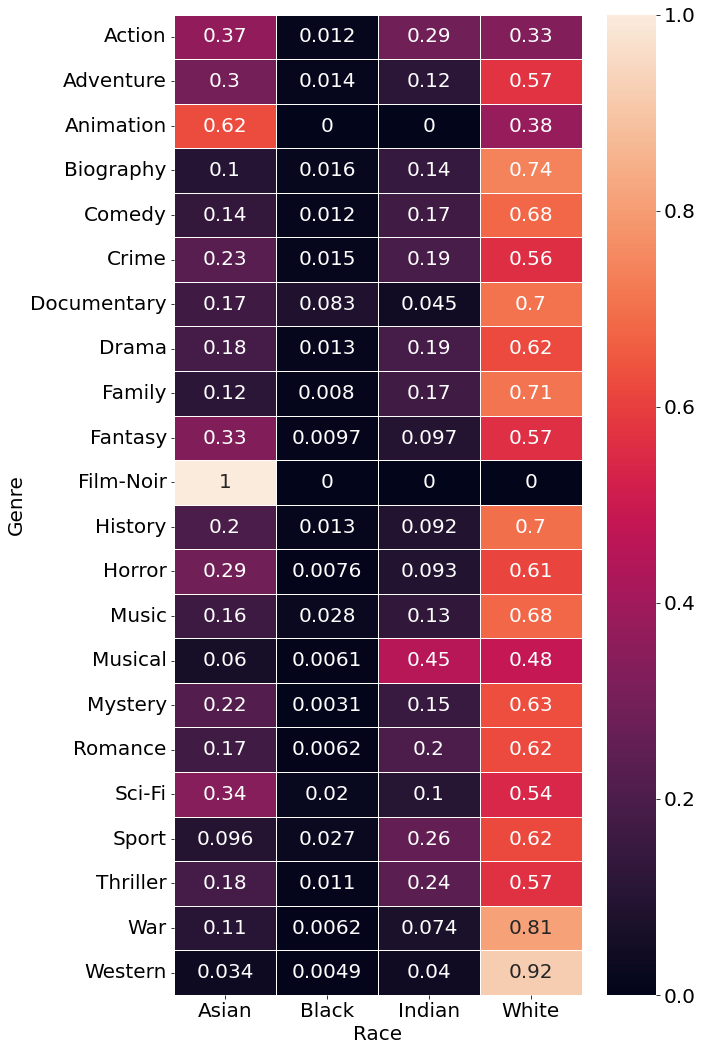}
    \caption{Racial distribution for each genre.}
  \end{subfigure}
     \hfill
     \begin{subfigure}[t]{0.49\textwidth}
         \centering
        \includegraphics[width=0.95\textwidth]{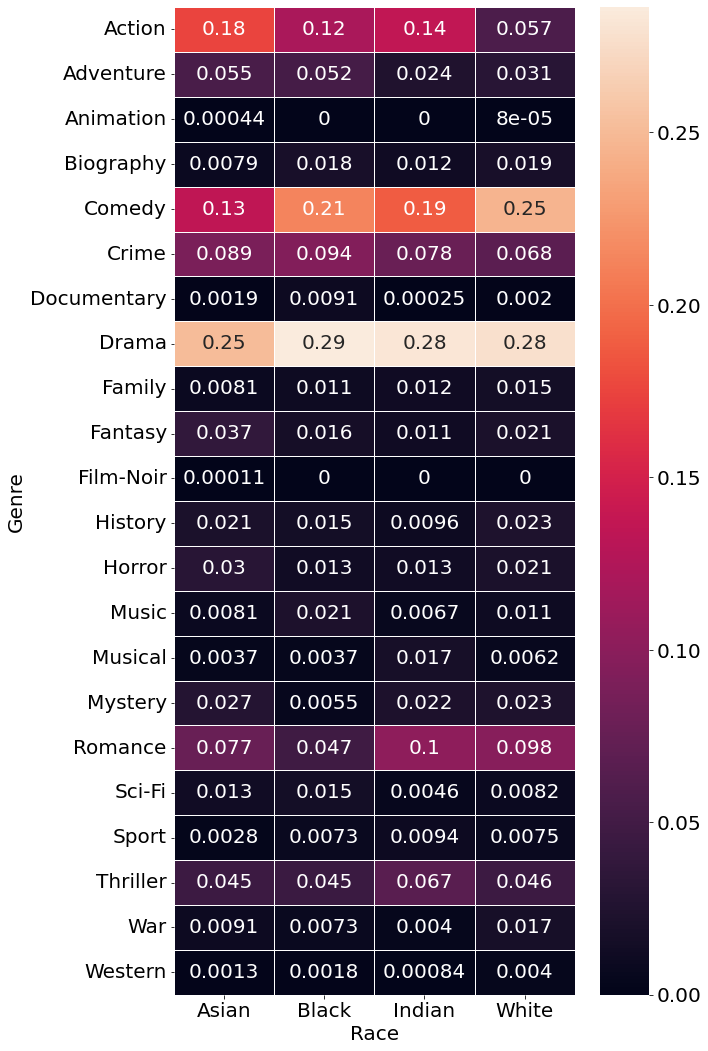}
    \caption{Distribution for each race of appearance on posters by genre. }
     \end{subfigure}
     \hfill
     \centering
     \caption{Heatmaps for probability of ethnic actors appearance in movie posters.}
\end{figure*}

\begin{figure}
        \includegraphics[width=\columnwidth]{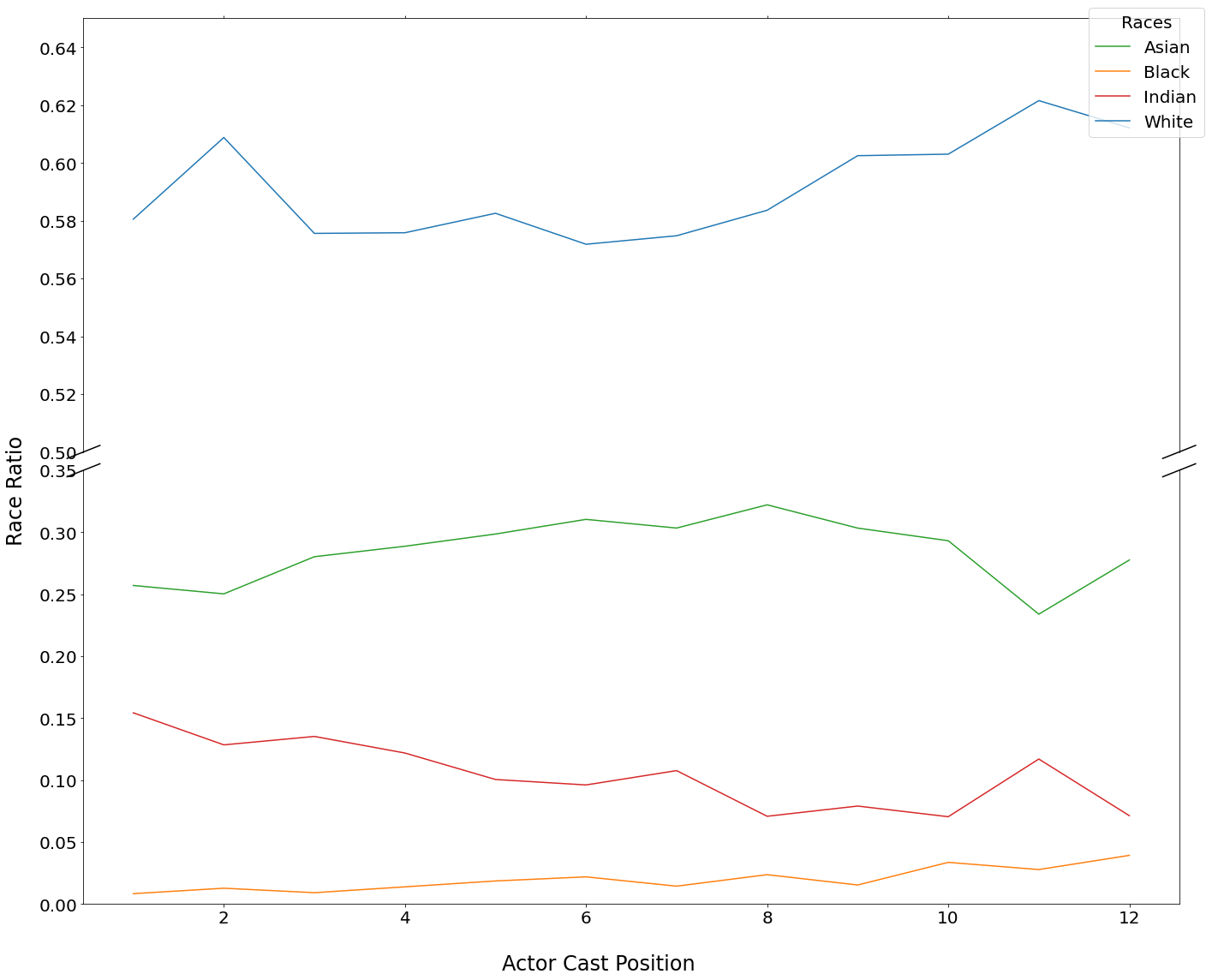}
     \hfill
     \centering
    \caption{Ethnic Related Appearance in each Rank Position: at each rank, the relative number of actors of each race category out of all actors is depicted.}
\end{figure}

\end{document}